\documentclass[]{aa}  
\usepackage{graphicx}
\usepackage{txfonts}
\usepackage{natbib} 
\bibpunct{(}{)}{;}{a}{}{,} 
\begin{document}
   \title{The circumbinary disc around the J-type C-star \object{IRAS\,18006-3213}\thanks{Based on observations collected at
   the European Southern Observatory, Chile. Program ID: 073.D-610(A).}}


\author{P. Deroo\inst{1}  \and  H. Van Winckel\inst{1} \and
  T. Verhoelst\inst{1} \and M. Min\inst{2}
  \and M. Reyniers\inst{1}  \and L.B.F.M. Waters\inst{1,2}}

\offprints{P. Deroo}

\institute{Instituut voor Sterrenkunde, K.U. Leuven, Celestijnenlaan
  200B, B-3001 Leuven, Belgium\\ \email{pieter.deroo@ster.kuleuven.be}
  \and Astronomical Institute ``Anton Pannekoek'', University of
  Amsterdam, Kruislaan 403, 1098 SJ Amsterdam, the Netherlands }

\date{}

 
  \abstract {In the generally accepted, but poorly documented model,
silicate J-type C-stars are binary objects for which the silicate
emission originates from a circumbinary or a circumcompanion disc.}
{We aim at testing this hypothesis by a thorough spectral and
spatial observational study of one object:
\object{IRAS\,18006-3213}. } {We obtained, analysed and modeled high
spatial resolution interferometric VLTI/MIDI observations on multiple
baselines ranging from 45\,m to 100\,m.} {All observations resolved
the object and show the very compact nature of the N-band emission
($\sim$ 30\,mas). In addition, the highest spatial resolution data
show a significant differential phase jump around 8.3\,$\mu$m. This
demonstrates the \emph{asymmetric} nature of the N-band
emission. Moreover, the single telescope N-band spectrum shows the
signature of highly processed silicate grains.  These data are used to
confirm the model on silicate J-type C-stars for
\object{IRAS\,18006-3213}. We show that the most favourable
model of the dust geometry is a stable circumbinary disc around the
system, seen under an intermediate inclination.}{ The data
presented on the silicate J-type C-star \object{IRAS\,18006-3213}
provide evidence that the oxygen rich dust is trapped in a
circumbinary disc. The formation of this disc is probably linked to
the binary nature of the central star.}


\maketitle
%

\section{Introduction}\label{sect:introduction_iras18006}

The $^{12}$C content in the atmosphere of asymptotic giant branch
(AGB) stars is believed to increase during the evolution along the
spectral sequence M $\rightarrow$ S $\rightarrow$ C. This increase is
the result of mixing of He-burning material (mainly $^{12}$C and
s-process elements) by the convective envelope through the third
dredge-up. This mixing process is thought to occur during the
relaxation after a thermal pulse. The recurrent operation of this
dredge-up leads to the creation of a carbon star, when carbon becomes
more abundant in number than oxygen.

The spectral classification of carbon stars \citep[][and references
  therein]{Wallerstein_1998} revealed, however, that different
  spectral groups could be identified and only the C(N) stars are now
  believed to be AGB stars, which are in the process of active
  dredge-up evolution. Other spectral classes are much less
  understood.

The J-type carbon stars are genuine carbon stars (C$>$O), but they
  show very low $^{12}$C/$^{13}$C values, at or near the CNO cycle
  equilibrium value of about 4. Moreover, J-type carbon stars show no
  s-process overabundances and often large Li abundances \citep[][and
  references therein]{Abia_2000}. All these chemical characteristics
  are at variance with the AGB chemical models and it is fair to say
  that the class of J-type C-stars are chemically far from
  understood. They do, however, represent a significant fraction of
  all C-stars ($\sim$ 10\,\%) as was shown from a recent survey of
  1497 stars in \cite{Morgan_2003}.

Because of the stability of the CO molecule, O-rich stars create
circumstellar envelopes with silicate dust emission, while C-rich
stars show C-rich dust features. The detection by IRAS/LRS of O-rich
dust around C-rich photospheres came therefore as a huge surprise
\citep{Little-Marenin_1986,Willems_1986}. Even more intriguing is the
observation that \emph{all} C-stars showing silicate dust are
spectroscopically of J-type \citep{LloydEvans_1991}. No photospheric
distinction can be made within the class of J-type C-stars, between
the silicate emitters and the others \citep{Ohnaka_1999}. The J-type
C-stars showing silicate dust emission are defined as ``silicate
J-type C-stars'' and constitute about 10\,\% \citep{LloydEvans_1991}
of the J-type C-stars.

Despite 20 years of research, the evolutionary channel(s) leading
to the formation of silicate J-type carbon stars remain(s) a mystery.
\cite{Little-Marenin_1986} suggested that silicate J-type C-stars can
be explained by assuming they are binaries consisting of a C-rich and
O-rich giant. However, no direct detection of this companion has ever
been made, neither spectroscopically \citep{Lambert_1990} nor with
speckle interferometry \citep{Engels_1994b}. \cite{Willems_1986} on
the other hand suggest that these objects experienced a recent
transition from O-rich to C-rich by means of a thermal pulse. The
O-rich matter in this hypothesis is then the remnant of a previous
mass-loss phase, when the giant was still O-rich. This was criticized
from an evolutionary point of view. \cite{LloydEvans_1990} argues that
such a transition object should only be observed during a few
decades, while some of the silicate C-stars are known to have C-rich
photospheres for over 50 years. Furthermore, it is shown in
\cite{Yamamura_2000} that the silicate feature of \object{V778\,Cyg}
did not change in the 14 years time interval between the IRAS/LRS and
the ISO/SWS observation. Therefore, also this scenario seems unlikely.

The general consensus now about these objects is that they are
binaries with an unseen companion, probably a main sequence star
\citep{Morris_1987, LloydEvans_1990, Yamamura_2000}. While the star
was an O-rich giant, the mass loss was captured and stored in a disc,
either around the companion or around the system. Subsequently, the
primary experienced (a) thermal pulse(s) and evolved into a
C-star. This scenario still does not account for
the chemical J-type nature of the silicate C-star.

Evidence of this hypothesis has been gathered over the years: The
long-lived reservoir has been inferred from very narrow rotational CO
line profiles by \cite{Kahane_1998} and \cite{Jura_1999}, indicative
of Keplerian rather than outflow velocities.  Radial velocity
measurements of two silicate C-stars were consistent with the motion
in a binary system. No conclusive orbit could be detected, however
\citep{Barnbaum_1991}. A recent observation of the water maser
emission toward \object{V778\,Cyg} at high angular resolution suggests
the presence of a rotating disc \citep{Szczerba_2006}. From the
angular separation between this emission and a previous optical
observation, the authors conclude that this (probably) provides
support for the circumcompanion model for this object
\citep{Yamamura_2000}.

Despite the more general consensus of the binary nature of silicate
J-type stars, conclusive direct evidence on their binarity is still
lacking.  It remains also unclear whether one formation channel does
applies for all silicate J-type C-stars. An open question is also
whether the scenario developed for the silicate J-type C-stars, could
apply for all J-type stars. Whether a possible binary nature can have
a direct impact on the internal chemical evolution of one of the
components is yet another open question.

To investigate the silicate J-type star problems, we studied
\object{IRAS 18006-3213} at high angular resolution using the MIDI
instrument.  \object{IRAS\,18006-3213} (\object{CGCS 3935},
\object{Fuen C 270}, \object{FJF 270}) was identified as a C-star more
than 25 years ago by \cite{Fuenmayor_1981}. Soon after, it was
realised by \cite{Willems_1986} that this object belongs to the
peculiar class of C-rich objects showing an O-rich circumstellar
environment. The J-type nature of the C-rich giant was confirmed by
\cite{LloydEvans_1990}. IRAS\,18006-3213 is therefore a perfect
candidate to provide answers to the questions posed above. The
distance to this object is uncertain and was quantified to be 2.6\,kpc
by \citet{Engels_1994} based on the matching between the measured
energy emitted in the infrared, and the assumption that the luminosity
is 5000\,L$_{\odot}$, which is typical for thermally pulsing AGB
stars.

The paper is organized as follows. In
sect.~\ref{sect:observations_and_reduction_iras18006} we present the
observations for which an analysis is given in
sect.~\ref{sect:global_analysis_iras18006}. The data provide
constraints on the source morphology which are discussed in
sect.~\ref{sect:discussion_iras18006}. The conclusions are given in
sect.~\ref{sect:conclusions_iras18006}.


\section{Observations and Reduction}\label{sect:observations_and_reduction_iras18006}

\subsection{observations}
Five interferometric observations of \object{IRAS\,18006-3213} were
performed using the MIDI instrument at different physical baselines
with projected baselines (PB) ranging from 40\,m to 102\,m. One of the
observations (with a PB of 40\,m) shows bad overlap
between both beams and is therefore discarded. A log of the successful
observations is presented in Table \ref{tab:log_iras18006}. Each
observation was performed using a prism (with spectral resolution
$\lambda / \Delta \lambda \sim 30$) producing spectrally dispersed
fringes over the N-band ($8 - 13\,\mu\rm{m}$).  The standard observing
sequence was used, for which we refer to \cite{Leinert_2004}.  To
correct for instrumental and atmospheric loss of coherence, a
calibrator of known diameter is measured using the same setup. The
visibility calibrators of the different science observations are given
in Table~\ref{tab:log_iras18006} as well.

\begin{table}
\caption{The log of the observations of \object{IRAS\,18006-3213}. For each
  observation, the corresponding visibility calibrator is given as well.}
\label{tab:log_iras18006}
\centering
\begin{tabular}{l l c r r r r}
\hline \hline
\multicolumn{1}{c}{night} & 
\multicolumn{1}{c}{telescope} &
\multicolumn{1}{c}{time} & 
\multicolumn{1}{c}{PB} &
\multicolumn{1}{c}{PA} &
\multicolumn{1}{c}{visibility} \\

\multicolumn{1}{c}{yyyy/mm/dd} &
\multicolumn{1}{c}{combination} &
\multicolumn{1}{c}{hh:mm} &
\multicolumn{1}{c}{(m)} &
\multicolumn{1}{c}{(\degr )} &
\multicolumn{1}{c}{calibrator} \\
\hline
2004-04-08 & UT2 - UT3 & 08:17 & 46.5  & 30 & HD\,165135\\
2004-06-27 & UT1 - UT3 & 03:15 & 102.4 & 25 & HD\,134505\\
2004-06-27 & UT1 - UT3 & 06:06 & 96.0  & 43 & HD\,139127\\
2004-06-28 & UT1 - UT3 & 08:48 & 70.8  & 49 & HD\,165135\\

\hline
\end{tabular}
\end{table}

\subsection{spectroscopic reduction}
\begin{figure}[t] 
\includegraphics[width=\hsize]{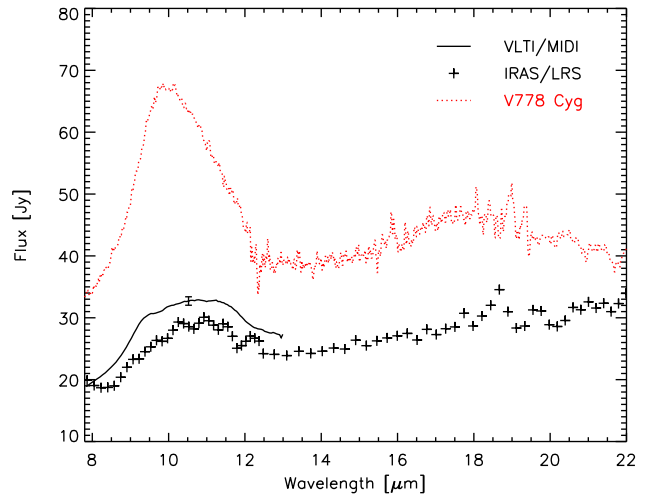}
\caption{Observed spectra of \object{IRAS\,18006-3213}. The MIDI
  spectrum is plotted using a solid line while the IRAS/LRS spectrum
  is shown using crosses. Clearly the shape in the N-band region is
  very similar between both spectra although the absolute level is not
  identical. Overplotted using a dotted line is the silicate J-type
  C-star \object{V778\,Cyg} which is shifted by 20\,Jy for the sake of
  clarity. The 10\,$\mu$m profile shows that the silicates around
  \object{IRAS\,18006-3213} are much more processed than the dust
  around \object{V778\,Cyg}.}\label{fig:spec_iras18006}\end{figure}

A traditional single telescope spectrum is obtained, together with
each interferometric observation. We corrected for atmospheric
transmission and obtained flux calibration using the photometric
observations of the calibrator stars. Theoretical N-band spectra of
the calibrator stars were synthesized using the {\sc marcs} code
\citep[][and further updates]{Gustafsson_1975}, including the SiO
opacity.  Temperature, gravity and angular diameter of the calibrators
were derived from SED fitting by \cite{Verhoelst_phd} (cf.
http://www.ster.kuleuven.ac.be/$\sim$tijl/MIDI\_calibration/mcc.txt).  The spectrum
of \object{IRAS\,18006-3213} obtained for 2004-04-08 is shown in
Fig.~\ref{fig:spec_iras18006}. Overplotted in this figure is the IRAS
low resolution spectrum scaled on the basis of the IRAS 12 $\mu$m
photometry flux. Also shown in this figure is the ISO/SWS spectrum of
\object{V778\,Cyg}.

\subsection{interferometric reduction}

To reduce the interferometric data, two different methods were
employed. The first method is based on power spectrum analysis and was
executed with the MIA package (MIDI interactive analysis, version 1.3)
implemented at the Max-Planck Institut f\"ur Astronomy in
Heidelberg. The second method reduces all frames to the same optical
path difference (OPD) and was performed using the EWS package (Expert
Work Station) developed by Walter Jaffe. For the details of both
reduction packages, we refer to \cite{Leinert_2004} and
\cite{Jaffe_2004} respectively.

\begin{figure}[t]
\includegraphics[width=\hsize]{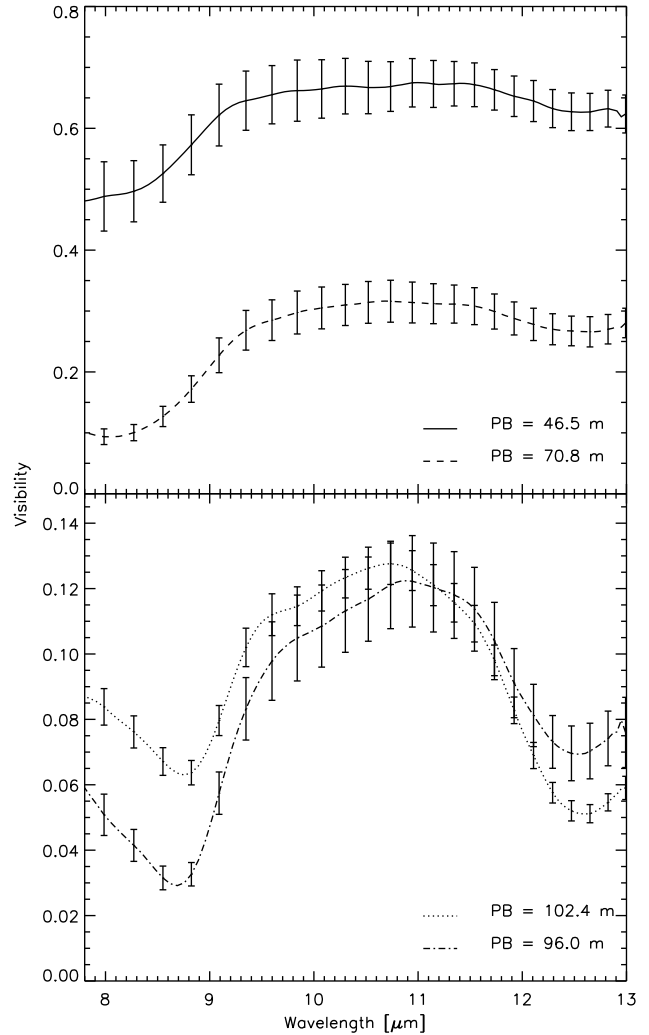}
\caption{The calibrated visibilities for the different measurements of
\object{IRAS\,18006-3213}. In the \emph{top panel} the visibilities
for the short baseline configurations are shown while in the
\emph{bottom panel} those for the longer baseline settings are
given. The error bars are the RMS value of the different scans as
estimated by the EWS
method.}\label{fig:visibilities_iras18006}\end{figure}

\begin{figure}[t]
\includegraphics[width=\hsize]{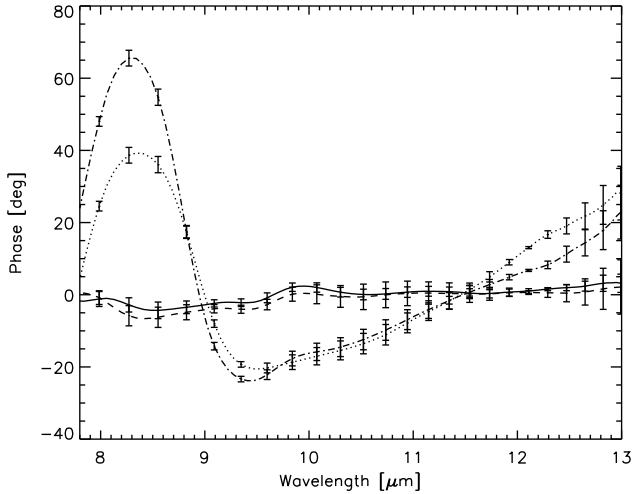}
\caption{The calibrated phase for the different measurements of
IRAS\,18006-3213. Note that only second order effects in the phase
variation over the wavelength band can be retrieved. The same line
styles were employed as the ones used in
Fig.~\ref{fig:visibilities_iras18006}. Note that for the two short
baseline measurements (solid line and dashed line) the phase is
centered around zero while for the long baseline measurements (dotted
and dashed-dotted line, a large phase jump is observed around 8.5
$\mu$m. The error bars are the RMS values on the instrumental phase
during the night. }\label{fig:phases_iras18006}\end{figure}

\begin{figure}[t]
\includegraphics[width=\hsize]{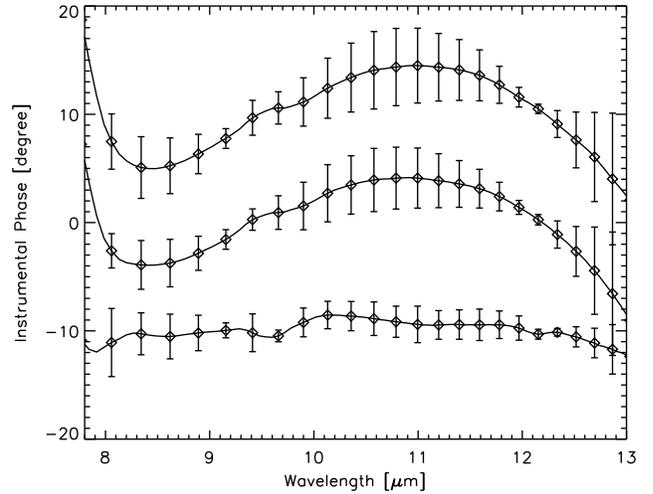}
\caption{The mean instrumental phase for the different observing
  nights. The curvature of the instrumental phase is caused by the
  varying values of index of refraction of air in the VLTI delay line
  tunnels. The error given is the RMS error on the different
  calibrator observations during the night. The data for 2004-04-08 is
  shifted down by 10 degree while the data for 2004-06-28 is shifted
  up by 10 degree for the sake of clarity. Clearly the atmospheric
  conditions were very stable during the observations giving rise to
  very small RMS errors of the order of only a few
  degree.}\label{fig:instrumental_phase_iras18006}
\end{figure}

In Fig.~\ref{fig:visibilities_iras18006} we show the visibilities
obtained with the EWS method. The error bars consists both of
systematic errors and noise type errors. Because the systematic
contribution to the errors is dominant, the actual point-to-point
error on the visibility is much smaller than indicated in
Fig. \ref{fig:visibilities_iras18006}, i.e. the relative visibility
between e.g. 8 and 9 $\mu$m is much better determined than the
numerical value at either wavelength. The visibilities obtained with
the MIA reduction scheme give very similar results. In the following,
we will work with the EWS reduced data because this  method has the
advantage that the relative phase between the different spectral
channels is preserved in the reduction. All first order information on
the phase delay is lost because it is necessary to compensate for OPD
fluctuations and first-order dispersion effects before co-adding the
data \citep{Jaffe_2004}. Second order effects remain in the data and
they can be determined very accurately. An accuracy of 1 degree RMS
phase error can be achieved during stable environmental conditions
\citep{Tubbs_2004}. In Fig.~\ref{fig:instrumental_phase_iras18006}, we
show the instrumental phase of the different observing nights as
measured for all calibrator objects observed during the night. The
instrumental phase is clearly very stable with an accuracy for all
nights of about 5 degree RMS phase error. The curvature in the
instrumental phase is a result of the varying values of the index of
refraction of air in the VLTI tunnels. The instrumental phase is
subtracted from the science target observations, largely removing this
curvature. As a conservative error on the relative phase, we took the
RMS phase degree error on all calibrator observations during the
night. The resulting phase information on IRAS\,18006-3213 for the
different observations is shown in Fig.~\ref{fig:phases_iras18006}.

\section{Analysis}\label{sect:global_analysis_iras18006}

\subsection{spectral energy
  distribution}\label{sect:spectral_energy_distribution}

\begin{figure}[t]
\includegraphics[width=\hsize]{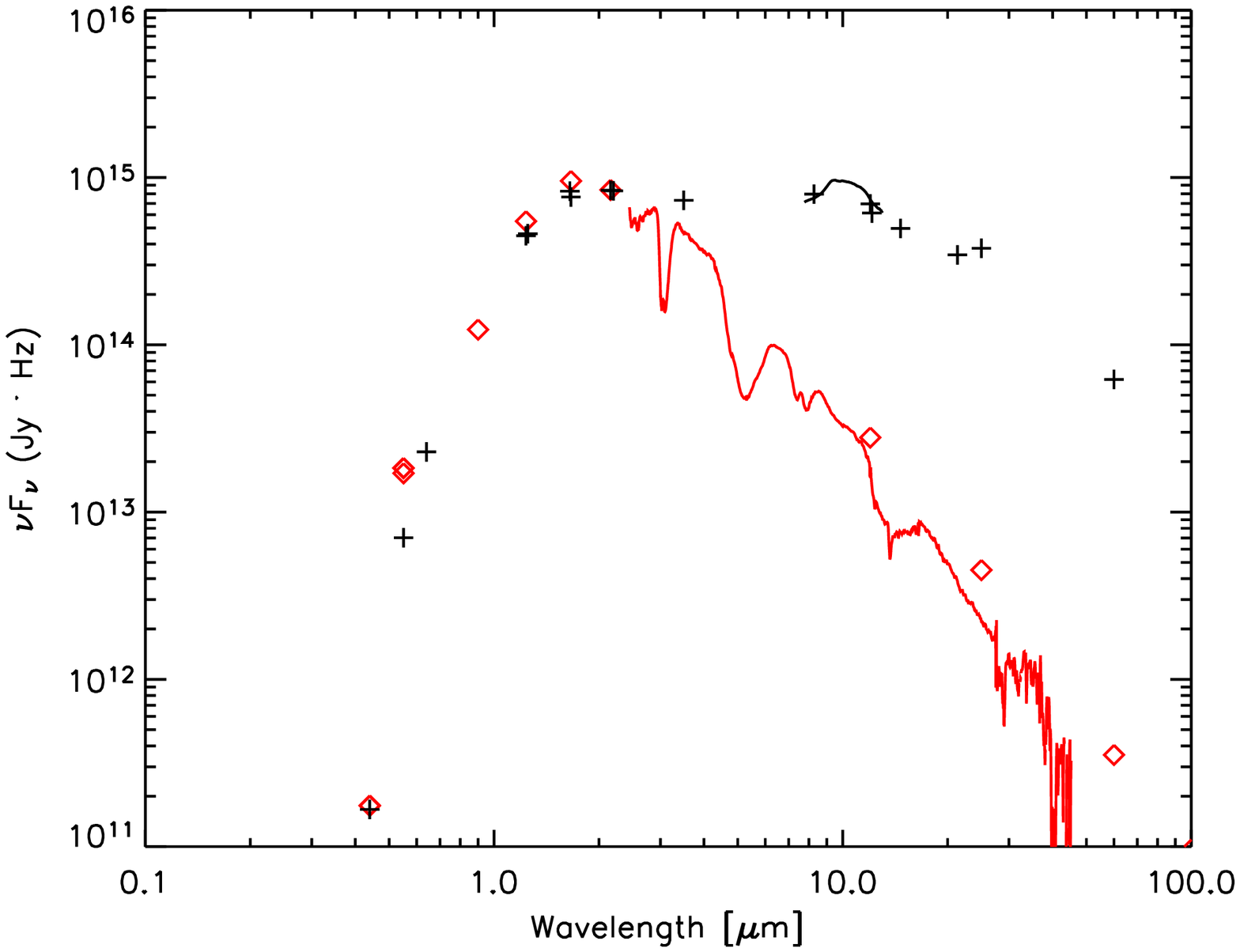}
\caption{The SED of \object{IRAS\,18006-3213} is shown using crosses
  while the SED of a comparable J-type C-star without silicate
  emission, \object{T\,Lyr}, is overplotted using diamonds. Shown
  using a solid line is the ISO/SWS spectrum of \object{T\,Lyr} and
  the MIDI spectrum of \object{IRAS\,18006-3213}. While both objects
  have similar colours in the optical, \object{IRAS\,18006-3213} shows
  a huge IR-excess in comparison to \object{T\,Lyr}.
  \object{IRAS\,18006-3213} can not be experiencing a large spherical
  C-rich outflow. }\label{fig:sed_iras18006}\end{figure}

To construct the spectral energy distribution (SED) of
\object{IRAS\,18006-3213}, we used optical photometry from the
literature \citep{LloydEvans_1990, LloydEvans_1991} and infrared data
from 2\,MASS, MSX and IRAS. The data are shown in
Fig.~\ref{fig:sed_iras18006} using crosses. The SED shows a clear
IR-excess representing an amount of energy almost equal to that
emitted in the optical.

We note that the DENIS database lists a K magnitude of almost 0.7
$\pm$ 0.1 mag lower than the one reported in the 2\,MASS database
whilst having compatible J magnitudes. This could indicate that the
(dust) excess emission in the infrared is variable. This is, however,
incompatible with both the spectral and photometric stability of the
mid-IR emission (see sect.~\ref{sect:mineralogy_iras18006}). In what
follows we did not take into account the DENIS K band magnitude.

In Fig.~\ref{fig:sed_iras18006}, the SED of
\object{IRAS\,18006-3213} is compared to the SED of \object{T\,Lyr}, a
J-type C-star of similar spectral type but without silicate emission.
Both objects are among the reddest members of the class
\citep{LloydEvans_1990, Lorenz_Martins_1996}. The spectral types
indicate low effective temperatures in the range of T$_{\rm{eff}} =
2000 - 2500$\,K. While both objects show comparable optical colours, in
the IR the SED is completely different. \object{IRAS18006-3213} shows
a large IR excess dominated by thermal emission of dust, while the IR
spectrum of \object{T\,Lyr} is dominated by molecular features and has
a very different spectral slope.  The total mass-loss of
\object{T\,Lyr} is quantified by \citet{Bergeat_2005} to be $8
. 10^{-7} \rm{M}_{\odot} \rm{yr} ^{-1}$ and there is no strong evidence
for efficient dust production in the wind.

Since both \object{T\,Lyr} and \object{IRAS\,18006-3213} have very
similar spectral type, and have very similar colours in the optical,
we can assume that also the photosphere of \object{IRAS\,18006-3213}
does not experience a large dust-driven spherical C-rich outflow and
that the IR spectrum of the underlying carbon star is very similar to
the template \object{T\,Lyr}.  In general, J-type objects show very
moderate outflows \citep{Lorenz_Martins_1996}.

We conclude that the spectral energy distribution of
\object{IRAS\,18006-3213} consists of a photospheric contribution and
a large infrared excess, likely completely dominated by oxygen rich
dust alone. 

\subsection{mineralogy}\label{sect:mineralogy_iras18006}

In Fig.~\ref{fig:spec_iras18006}, the MIDI spectrum of
\object{IRAS\,18006-3213} is shown together with its IRAS/LRS
spectrum. Both show good agreement, which indicates that the silicate
emission must have been stable for over 20 years. This is consistent
with the stable emission found for the two other silicate J-type
C-stars of which the IR excess was observed at different epochs
\citep[\object{V778\,Cyg} and
\object{Hen\,38},][respectively]{Yamamura_2000,Ohnaka_2006}.

The band profile on the N-band silicate feature is a good tracer
for the mineralogy of the dust emission.  Comparing the spectrum of
\object{IRAS\,18006-3213} with that of \object{V778\,Cyg} (also shown
in Fig.~\ref{fig:spec_iras18006}), another silicate J-type C-star, it
is striking that the silicate emission profiles are very
different. The profile of \object{IRAS\,18006-3213} has the signature
of highly processed silicates (i.e. crystalline and/or large grains)
while \object{V778\,Cyg} shows evidence for an unprocessed silicate
environment (i.e. amorphous small grains).

In order to quantify the difference in mineralogy and sizes of
the emitting dust particles around both objects, we compute a fit through the
N-band spectra using calculated emissivities of irregularly shaped,
chemically homogeneous dust grains. We ignored the contribution
provided by their photospheres because this contribution and the
effect on the slope of the N-band  spectrum is
marginal. The most important dust species causing spectral signature
in the 10\,$\mu$m window are amorphous (i.e. glass with olivine
stoichiometry) and crystalline olivine, amorphous (i.e. glass
with pyroxene stoichiometry) and crystalline pyroxene and amorphous
silica. In the fitting, we added also a continuum distribution which
accounts for the emission by large grains and/or for the possible
presence of featureless components such as metallic iron, iron sulphide
or carbon grains.  This continuum contribution is modeled using a
constant mass absorption coefficient. For a reference to the method
which is used and the assumptions made, we refer to \cite{Min_2005}
and \cite{Vanboekel_2005}.

In the 10\,$\mu$m region we are mainly sensitive to the dust
grains smaller than a few $\mu$m. We represented the size distribution
of the particles by two grain sizes, 0.1\,$\mu$m and 1.5\,$\mu$m.
This size is the radius of the solid sphere with an equivalent volume
as the hollow sphere \citep{Min_2005}. Particles larger than a few
$\mu$m contribute mainly to the continuum. In addition, we assumed
that the thermal radiation we analyze originates from optically thin
regions so that we can simply add the different contributors to the
spectral emission feature.

In the fitting process of the N-band spectrum, we assume all
contributing dust grains, including the ones responsible for the
continuum, to have the same temperature distribution. Due to the
limited wavelength range, this temperature distribution can be
represented by a single Planck curve with one characteristic
temperature, $T_{\rm{c}}$. This is justified because it is likely that
the dust grains of different species are coagulated, implying thermal
contact between the various components. The characteristic
temperatures used in the modeling are given in
Table~\ref{tab:chemistry_iras18006}. Note that this temperature can
not be regarded as the temperature of the whole dust excess but is
only an approximation of the temperature of those particles
responsible for the N-band emission.

The abundances of the dust components are determined by using a
linear least square fitting procedure with constraints on the weights
to avoid negative values. The temperature of the grains and the
underlying continuum is varied from 0 to 1500~K until a best fit is
obtained. The errors on the resulting fit parameters are determined
through a Monte Carlo simulation on the spectrum. This method ensures
that degeneracies between the fit parameters show up as large errors
in those parameters. 

\begin{figure}[t]
\includegraphics[width=\hsize]{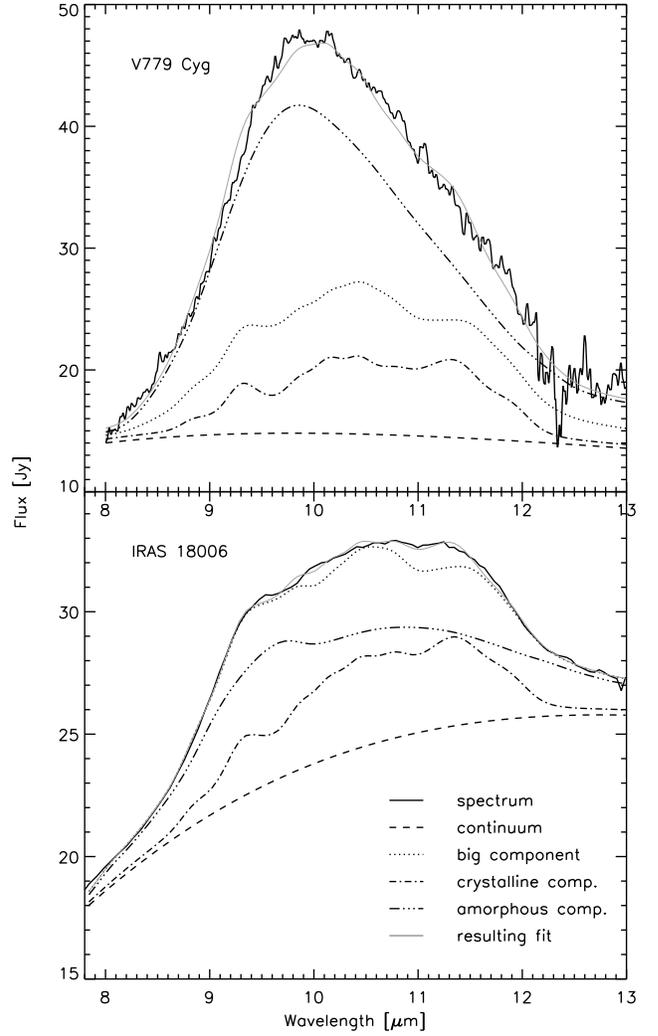}
\caption{The N-band spectra of \object{V778\,Cyg} (top) and
  \object{IRAS\,18006-3213} (bottom) are shown using a black solid
  line. Overplotted using a grey line are the best fit models. Note
  the almost perfect agreement between the fit and the spectrum. Also
  shown in the figure are: the continuum contribution, the emission
  produced by 1.5\,$\mu$m sized particles (i.e. the big component),
  the crystalline component and the amorphous component. The dust
  around both objects is clearly very different. While the spectrum of
  \object{V778\,Cyg} is reproduced mostly through features associated
  with small grains, the circumstellar environment of
  \object{IRAS\,18006-3213} consists almost entirely of large
  (processed) grains.}\label{fig:chemistry_iras18006}
\end{figure}

The dust parameters and its respective errors for
\object{IRAS\,18006-3213} and \object{V778\,Cyg} are given in
Table~\ref{tab:chemistry_iras18006}. The resulting best fit spectrum
is shown as a grey line in
Fig.~\ref{fig:chemistry_iras18006}.

\begin{table}
\caption{The composition and grain sizes of the silicate dust in the
  circumstellar environment around \object{IRAS\,18006-3213}
  (IRAS\,18006) and \object{V778\,Cyg}. The percentages for the
  different particles are listed in mass fraction. The olivine and
  pyroxene particles are amorphous while forsterite and enstatite are
  crystalline particles. Two grain sizes were adopted in the fit:
  small particles are 0.1\,$\mu$m while big particles are 1.5\,$\mu$m
  large. Also given is the total amount of big particles needed in the
  fit and the overall crystallinity fraction. Clearly the dust around
  \object{IRAS\,18006-3213} is much more processed as
  \object{V778\,Cyg}, both in grain size and crystallinity.}
\label{tab:chemistry_iras18006}
\centering
\begin{tabular}{l c c c}
\hline \hline
\multicolumn{1}{c}{} & 
\multicolumn{1}{c}{unit} &
\multicolumn{1}{c}{value} &
\multicolumn{1}{c}{value} \\
\multicolumn{1}{c}{} & 
\multicolumn{1}{c}{} &
\multicolumn{1}{c}{IRAS\,18006} &
\multicolumn{1}{c}{V778\,Cyg} \\
\hline
$T_{\rm{c}}$ & K & 400 $\pm$ 100 & 500 $\pm$ 100\\
small olivine & \% & 0.1 $\pm$ 0.1 & 61.9 $\pm$ 3.9\\
big olivine & \% & 57.4 $\pm$ 1.8  & 20.6 $\pm$ 6.4\\
small pyroxene & \% & 0 $\pm$ 0 & 0 $\pm$ 0\\
big pyroxene & \% &  0 $\pm$ 0 & 0.1 $\pm$ 0.1\\
small forsterite & \% &  3.1 $\pm$ 0.3 & 1.9 $\pm$ 0.3 \\
big forsterite & \% & 15.5 $\pm$ 0.9 & 9.3 $\pm$ 1.2\\
small forsterite & \% & 0 $\pm$ 0 & 0 $\pm$ 0\\
big enstatite & \% & 15.7 $\pm$ 1.1 & 6.3 $\pm$ 1.4\\
small silica & \% & 0 $\pm$ 0 & 0 $\pm$ 0\\
big silica & \% & 8.1 $\pm$ 0.3 & 0 $\pm$ 0\\
\hline
\multicolumn{3}{c}{summary} \\
\hline
big components & \% & 96.8 $\pm$ 0.5 & 36.2 $\pm$ 4.4\\
crystalline components & \% & 34 $\pm$ 2 & 17.4 $\pm$ 2.4\\
\hline
\end{tabular}
\end{table}

The circumstellar silicates around both objects are clearly very
different. The warm silicates of \object{IRAS\,18006-3213} are much
more processed than those of \object{V778\,Cyg}. While the fraction of
large grains is almost 100\,\% (their contribution to the spectrum is
shown using a dotted line in Fig.~\ref{fig:chemistry_iras18006}), they
represent only one third of the mass for \object{V778\,Cyg}. Moreover,
also the crystallinity fraction for \object{IRAS\,18006-3213} is twice
as large as the one determined for \object{V778\,Cyg}.

The presence of the high processing observed for
\object{IRAS\,18006-3213}, both with respect to grain size and
crystallinity degree, is indicative of dust stored in a stable disc
\citep[e.g.][]{Molster_2000, Vanboekel_2005}. In O-rich outflows
observed around mass-losing objects, the dust is mostly in the form of
amorphous silicates \citep[e.g.][]{Sloan_1995, Waters_1996L,
Cami_2002}. The spectrum of \object{V778\,Cyg} is compatible with the
emission typically observed in such outflows. We can conclude that the
degree of processing of the circumstellar environment of silicate
J-type C-stars is clearly not uniform. 
%
%

\subsection{N-band visibilities}

In this section we discuss the global characteristics of the N-band
emission as constrained by the visibilities shown in
Fig.~\ref{fig:visibilities_iras18006}. The object is clearly resolved
at all baselines. In addition, a strong wavelength dependence is
observed in the visibility observations.

To quantify the global characteristics of the N-band emission, we
modeled the visibility measurements assuming using a simple geometry,
i.e. a uniform disc intensity distribution. We limit this part of the
analysis to the two visibility curves at the lowest spatial frequency,
because only these probe the large scale structure. The highest
spatial frequency data yield very low visibilities and their
interpretation is therefore very sensitive to the exact unresolved
photospheric contribution. The possible presence of small scale
structures in the dust excess like clumps, warped structure of the
disc, etc.. are very poorly constrained with the limited number of
visibilities observed. The long baseline measurements are, however,
used for their phase information, see
sect. \ref{sect:N-band_phases_iras18006}.

\begin{figure}[t]
\includegraphics[width=\hsize]{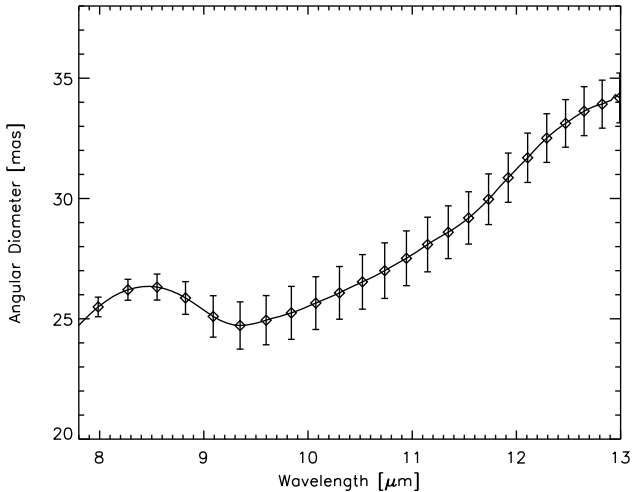}
\caption{The angular size of the emission in the N-band as determined
  using a uniform disc model. The fit was performed using the two
  visibility measurements at the shortest baselines only and should
  therefore only be considered as a rough first order
  estimate. Note that the point-to-point error is much smaller than
  the error given in the plot. }\label{fig:UDfit_iras18006}
\end{figure}

The resulting angular diameter of the N-band emission is shown in
Fig.~\ref{fig:UDfit_iras18006}. The dust emission is clearly compact
(25\,mas at 8\,$\mu$m to 35\,mas at 13\,$\mu$m) and it shows a clear
wavelength dependence. Globally, the size increases as wavelength
increases. This is likely the result of a temperature distribution,
with the colder dust located in a larger structure than the hotter
dust. In addition, a clear dip in angular size is observed starting
around 9\,$\mu$m.  Since the error bars in
Fig.~\ref{fig:UDfit_iras18006} represent mostly a scaling uncertainty
in the fluxes, the relative difference observed between e.g. 8.5\,$\mu$m and
9.5\,$\mu$m is significant.  This dip in the derived angular size
is the result of the shape of the visibility curve, which increases
rather steeply around 9 $\mu$m and flattens out till 11.5 $\mu$m after
which the visibility decreases (see top panel in
Fig.~\ref{fig:visibilities_iras18006}). Note that a Gaussian intensity
profile provides the same qualitative results.


Given the amount of emission and the size of the emission region at
8\,$\mu$m, which originates from continuum emitting particles (see
Fig.~\ref{fig:chemistry_iras18006}), it is clear that the
\emph{continuum emission must come from an optically thick region},
i.e. $\tau > 1$: If the emission were to come from an optically thin
region, the derived temperature would be inconsistent with the
SED. For example, if $\tau = 0.1$, then a temperature $> 1200$\,K is
necessary to reproduce the flux at 8\,$\mu$m.

\object{IRAS\,18006-3213} was identified as a C-star by
\cite{Fuenmayor_1981}. Therefore, any O-rich dust present in the
circumstellar environment must have been ejected by the AGB giant at
least 25\,years ago. Assuming a typical outflow velocity of
10\,km\,s$^{-1}$ and a distance for \object{IRAS\,18006-3213} of
2.6\,kpc \citep{Engels_1994}, the O-rich dust would need
$\sim$~15\,years to arrive at the derived radius of the dust
emission. This shows that the current size of the disc is in
contradiction with an outflow scenario, since the actual central
carbon star will not produce silicate dust.

The fact that the ``feature'' in the visibility profile has the same
shape as the silicate emission profile (see
Fig.~\ref{fig:spec_iras18006}) leads us to believe that the dip in the
angular size comes from a silicate emission feature originating from a
smaller radius than the underlying continuum emission.
\cite{Ohnaka_2006} observes a similar trend for the silicate carbon
star \object{IRAS\,08002-3803} and they show from a careful modeling
that a two-species model can reproduce the observations. In their
model, the silicate component has a steeper temperature gradient than
the other component due to the different opacities of both
particles. Therefore, the emission from the silicate component
originates from a smaller angular size. However, this is not the only
possible interpretation. An alternative explanation for the observed
``feature'' in the visibility curve is that different grain types do have
a different spatial density distributions. This was shown already in
some discs around Herbig Ae stars \citep{Vanboekel_2004nature} and for
the disc around the binary post-AGB object HD\,52961
\citep{Deroo_2006}.

In summary, the N-band interferometric observations clearly resolved
the circumstellar environment of \object{IRAS\,18006-3213} and
revealed the compact nature of the emission region. The size of the
emission provides strong indication that the N-band
emission must come from a compact stable geometry in the system.  Moreover,
the observations show that the particles responsible for the silicate
emission feature must be emitting from a smaller region than the
particles responsible for the continuum emission.

\subsection{N-band phases}\label{sect:N-band_phases_iras18006}

The calibrated phases (Fig.~\ref{fig:phases_iras18006}) show a large
phase-jump around 8.3\,$\mu$m for the two observations at the longest
baseline. This phase jump does not equal 180\,deg and/or it is not
accompanied with a visibility transition through zero\footnote{Note
that, due to the finite bandpass of MIDI, measuring a visibility which
equals zero is impossible. In reality, a monochromatic zero visibility
would correspond in our polychromatic measurement with V$\sim 0.008$,
which is clearly not observed in the measurements.}. Moreover, the
smooth shape of the phase curvature indicates that it is a slowly
evolving feature. This phase information shows that the source is
\emph{not} centrally symmetric with an identical photocenter
throughout the N-band.

The photospheric contribution at 10\,$\mu$m of the total received
  flux is 5\,$\pm$\,1\,\% (assuming a blackbody with the expected
  T$_{\rm eff}$, see sect.~\ref{sect:spectral_energy_distribution}). 
At the longest baselines, the dust emission is strongly
  resolved. The unresolved photosphere therefore becomes a significant
  contributor to the correlated flux of the system. The asymmetry in
  the system is therefore most likely explained assuming an offset
  between the photocenter of the dust excess and the giant
  photosphere. 

A photocenter offset between both components (the unresolved
  photosphere and the resolved circumstellar dust emission) results
  in the observed phase through the combination of the following
  effects:

\begin{itemize}
  \item The components, with different positions on the sky, have
  different spectral energy distributions. Therefore, the flux of both
  sources will contribute different fractions of the total received
  flux at different wavelengths, producing a phase difference
  between the spectral channels.
  \item The components have different positions on the sky and this
  displacement is more resolved at 8 $\mu$m than at 13 $\mu$m, leading
  to a modulation of the phase signal with the spatial resolution. 
\end{itemize}

The probability that the phase-jump is explained solely by the second
effect is very small, since the same phase-jump is observed at
different instrumental setups (see
Fig.~\ref{fig:phases_iras18006}). Indeed, finding the same shape in
phase with different projected baseline lengths and angles can only be
understood if we assume a very specific source geometry: the projected
angle and projected asymmetry of the source on the sky must be such
that they cancel in the different observed setups. The probability of
this geometry is clearly extremely low.

To see how the shape of the phase over the N-band can be reproduced by
different spectral signatures of both components (in combination with
the changing spatial resolution over the wavelength band), we
quantified a simple illustrative model.

The most important non-linear feature in the phase is centered around
8.3\,$\mu$m (see Fig.~\ref{fig:phases_iras18006}). For the
observations at the longest baselines, a strong positive slope is
observed before 8.3~$\mu$m. Between 8.3~$\mu$m and 9.3~$\mu$m the
phase strongly declines. In our model, an absorption/emission feature
in the giant photosphere and/or the dust spectrum could cause these
slopes. The most straightforward identification is the silicate
feature in the dust emission itself. This feature changes the flux
ratio drastically between 8.3\,$\mu$m and 9.3\,$\mu$m (see
Fig.~\ref{fig:chemistry_iras18006}). While the emission from the dust
excess increases by almost 50\,\% over this spectral region, the
emission of the giant photosphere decreases by 20\,\% in the
assumption of a Rayleigh-Jeans law. Clearly, the global photocenter
will shift rapidly toward the photocenter of the dust emission moving
from 8.3\,$\mu$m to 9.3\,$\mu$m. In the spectral region before
8.3\,$\mu$m, the presence of a strong absorption/emission feature in
the dust spectrum is excluded. The unresolved photosphere is the only
component for which a large change in flux contribution over that
specific spectral region would go unnoticed in the global N-band
spectrum. The most likely source is the presence of a molecular
photospheric absorption band due to C$_2$H$_2$ and HCN located around
7.5\,$\mu$m. This feature appears to be omnipresent in C-stars as is
demonstrated by ISO \citep{Aoki_1999, Jorgensen_2000, Yang_2004,
  Gautschy-Loidl_2004} and Spitzer \citep{Matsuura_2006, Sloan_2006,
  Zijlstra_2006}. As a template of what this molecular band for
\object{IRAS\,18006-3213} could look like, we took the ISO/SWS
spectrum of \object{T\,Lyr}, a J-type C-star of similar spectral type
but without silicate dust. Its photospheric molecular band at
7.5\,$\mu$m is likely somewhat polluted by its moderate mass loss, but
the band shape is expected to remain the same. Moreover, the
double-peaked shape of the band of \object{T\,Lyr} (see
Fig.~\ref{fig:sed_iras18006}) is representative for the typical band
shape due to C$_2$H$_2$ and HCN: see e.g. the left panel of Fig.\,11
in \cite{Lagadec_2006}, where the averaged spectra around the
7.5\,$\mu$m band of 14 SMC C-stars is shown.

\begin{figure}[t]
\includegraphics[width=\hsize]{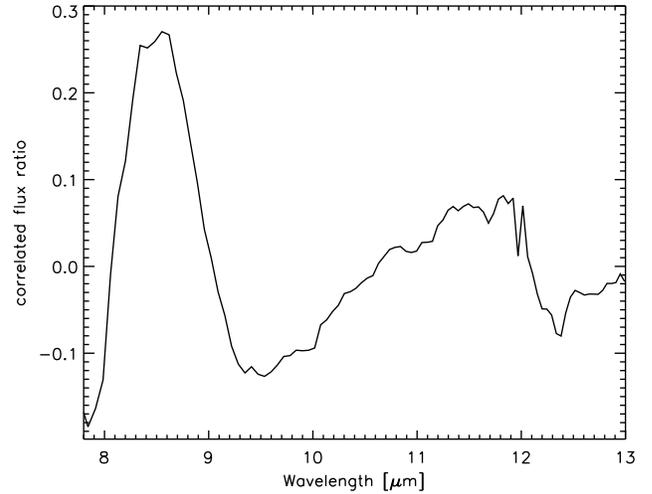}
\caption{A model for the correlated flux ratio between the infrared
  excess and the photospheric emission of
  \object{IRAS\,18006-3213}. For the infrared excess we assumed a
  visibility of 5\,\% over the wavelength band and for the
  photospheric emission we assumed the photospheric emission of
  \object{T\,Lyr} as reference. To be able to compare with the
  observed phase behaviour over the wavelength band (see
  Fig. \ref{fig:phases_iras18006}), we removed the first order
  behaviour of this
  ratio.}\label{fig:correlated_flux_ratio_iras18006}
\end{figure}

In Fig. \ref{fig:correlated_flux_ratio_iras18006} we show the
correlated flux ratio between both components assuming an unresolved
photosphere and a constant visibility over the wavelength band for the
dust excess. In order to compare directly the wavelength dependence of
this ratio with the observed phase (see
Fig. \ref{fig:phases_iras18006}), we removed its first order behaviour
over the wavelength band in
Fig. \ref{fig:correlated_flux_ratio_iras18006}. The observed phase
shape over the N-band is reproduced well with the phase jump located
at the observed wavelength position. The strength of the phase jump
around 8.3\,$\mu$m is enhanced if we assume the infrared dust excess
to have a lower visibility. The shape of the phase jump observed at
the longest baselines is therefore explained very well by: A shift of
the global photocenter toward the giant photosphere in the region
between 7.8\,$\mu$m and 8.3\,$\mu$m, followed by a shift toward the
photocenter of the dust excess between 8.3\,$\mu$m and 9.3\,$\mu$m.


The low visibilities at the longest baselines are determined by a
complex addition of an unresolved photosphere and a circumstellar dust
component with a different phase. This dust component has an
unconstrained (inner) geometry.  To provide nevertheless a quantified
estimate of the offset between both components, we assumed in our
model the following: {\it (i)} The emission of the C-giant can be
represented by the photospheric emission of \object{T\,Lyr}. {\it
(ii)} We assume that the wavelength dependence of the diameter of the
dust excess obtained at the shorter baselines (see
Fig.~\ref{fig:UDfit_iras18006}) is also applicable at the longest
baselines.

\begin{figure}[t]
\includegraphics[width=\hsize]{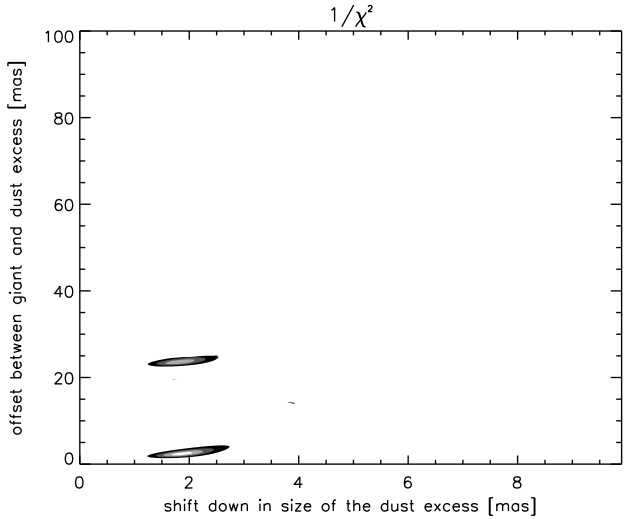}
\caption{The inverse $\chi^2$ distribution of the minimalisation
  between our model and the observations at the 96.0\,m baseline. Four
  contours are shown ranging from 0.15 to 0.30 in steps of 0.05. Two
  solutions are easily identified with distances between the giant
  photosphere and the dust excess of 2.4\,mas and 23.5\,mas. The
  optimum solution is compared to the observations in
  Fig.~\ref{fig:optimumminimalisation_iras18006}.}\label{fig:inversechisquared_iras18006}
\end{figure}

\begin{figure}[t]
\includegraphics[width=\hsize]{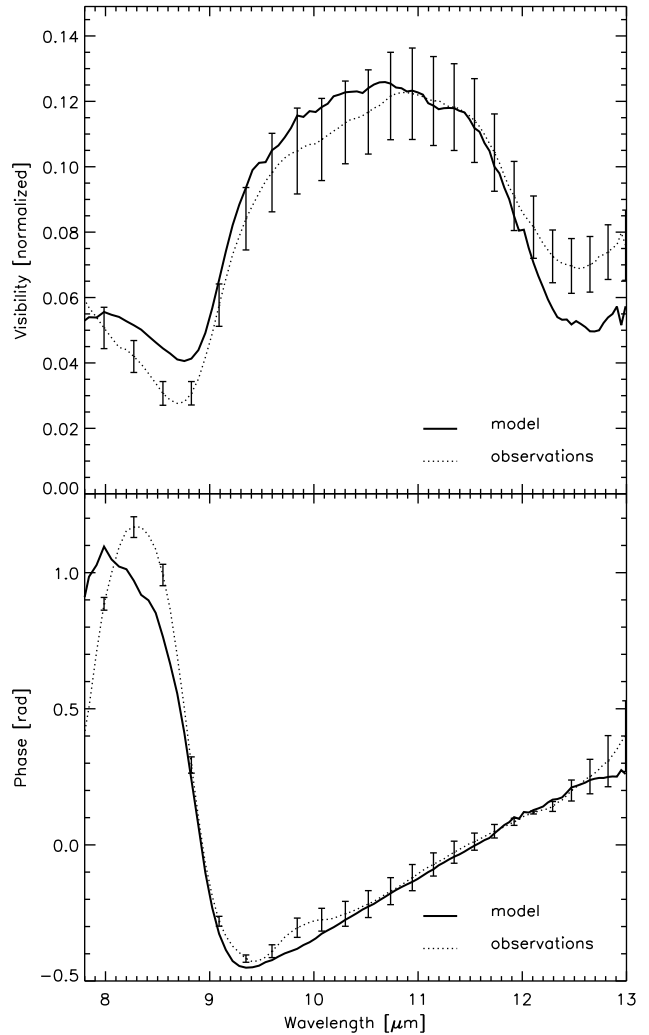}
\caption{The observed visibilities and phases at the 96.0\,m baseline
  are compared to the best fit model with a separation of 2.4\,mas
  between the giant photosphere and the photocenter of the dust
  excess. The model reproduces the characteristics of the data very
  well. Note that the worst part of the fit is located before
  8.3\,micron, where the effect of the uncertainty in the shape and
  strength of the molecular band in the giant photosphere is
  large. }\label{fig:optimumminimalisation_iras18006}
\end{figure}

Given the previous considerations and assumptions, we modeled the
visibility and phase observed at the 96.0\,m baseline using two free
parameters. The first parameter is a shift down in the size of the
dust excess as determined from the measurements at the shortest
baselines (see Fig.~\ref{fig:UDfit_iras18006}), to simulate
geometrical effects at high spatial resolution. The second parameter
is the offset between the giant photosphere and the photocenter of the
dust excess. In Fig.~\ref{fig:inversechisquared_iras18006}, the
inverse $\chi^2$ distribution is shown. Two solutions are found: The
optimum solution provides a separation between both components of only
2.4\,mas ($\chi^2 = 3.1$), while the other provides a much larger
separation of 23.5\,mas ($\chi^2 = 3.5$). The resulting fit to the
visibility and phase for the optimum solution is shown in
Fig.~\ref{fig:optimumminimalisation_iras18006}. Both observables are
reproduced very well considering the simplicity of the model. Modeling
the measurement at the 102.4\,m baseline provides similar results with
best fit separations between both components of 4.7\,mas and
14.9\,mas. We remark that under the assumptions we made, the
visibilities confirm the non-zero phase which is measured. 
To model the shape and the absolute value of the visibility amplitude, 
a complex (i.e. non-real) addition of the visibilities is necessary.


\section{Discussion}\label{sect:discussion_iras18006}

\subsection{location of the dust}\label{sect:location_of_the_dust_iras18006}

Silicate C-stars are identified as objects displaying a C-rich
photosphere with an O-rich circumstellar environment. Within all
the known silicate C-stars, large differences in silicate mineralogy
are observed. This was demonstrated in
sect.~\ref{sect:mineralogy_iras18006} and apparent from the wide
variety in 10\,$\mu$m silicate profiles \citep[see e.g. Fig. 4 in
][]{LloydEvans_1990} for silicate J-type C-stars.  While some objects
show a mineralogy which resembles the mineralogy observed in O-rich
AGB outflows (e.g. \object{V778\,Cyg} and \object{BM\,Gem}), other
show signatures of highly processed grains
(e.g. \object{IRAS\,08577-6035} and \object{EU\,And}). The object
which is at the focus of this study, \object{IRAS\,18006-3213},
clearly belongs to the latter class and shows a crystallinity degree
of 35\,\% and a fraction of large micron sized particles of almost
100\,\%.

In \cite{Yamamura_2000}, it is shown that the geometry which best
reproduces the ISO/SWS observation of \object{V778\,Cyg}, consists of
an O-rich reservoir located around the companion star. Because O-rich
AGB stars typically show amorphous outflows, the material captured by
the component is likely dominated by amorphous grains. Once in the
disc, part of the material in the inner region close to the low
luminosity companion, may be heated above the glass temperature at
which point it will become crystalline. The processed grains will
likely not provide a strong signature in the integrated spectrum and
the spectrum will be dominated by the evaporation of the amorphous
outer layers of this circum-companion disc. This evaporation is
triggered by the strong radiation pressure of the primary AGB giant.

In this scenario it is, however, difficult to account for the high
processing degree observed for the circumstellar environment of
\object{IRAS\,18006-3213}. The alternative geometry is when the
circumstellar material is captured in a circumbinary disc. In this
geometry, the inner material facing the central binary, is heated
above the glass temperature and can become crystalline. Because of the
high-luminosity AGB giant, this region is significantly larger. The
impact of this material on the spectrum is therefore expected to be
significant, thus providing an overall processed feature.

The visibility measurements of \object{IRAS\,18006-3213} also favour a
geometry where the dust is stored in a circumbinary disc. If the
silicate emission were to come from evaporation of the amorphous outer
layers of the disc, this emission would be larger than the underlying
continuum. Instead, the opposite is observed for
\object{IRAS\,18006-3213}.

\subsection{asymmetry in the system}

As shown in Sect. \ref{sect:N-band_phases_iras18006}, asymmetry is
present in the source geometry. This asymmetry is interpreted as
originating from a photocenter offset between the contribution of the
unresolved C-rich AGB photosphere and the resolved thermal emission of
circumstellar dust.  An offset between both components can be a
natural result of either:

\begin{itemize}
\item the motion of the AGB giant in its (suspected) binary orbit
\item the presence of a circumbinary disc seen under a non pole-on
inclination
\item the presence of a disc around the companion
\end{itemize}

%

While these scenarios provide different physical offsets between
the giant and the dust excess, the expected asymmetries are compatible
with the solutions found in sect.~\ref{sect:N-band_phases_iras18006}.
The MIDI spectrum and the visibility amplitudes however point to
the occurrence of a circumbinary disc rather than a circumcompanion
disc. We therefore evaluate one of the first two options more likely.


We conclude that -- although the interferometric phase
information is not conclusive -- an inclined circumbinary disc is the
most probable source of asymmetry in the system, possibly in
combination with the binary motion of the AGB giant.

\subsection{comparison with binary post-AGB objects}

\begin{figure}[t]
\includegraphics[width=\hsize]{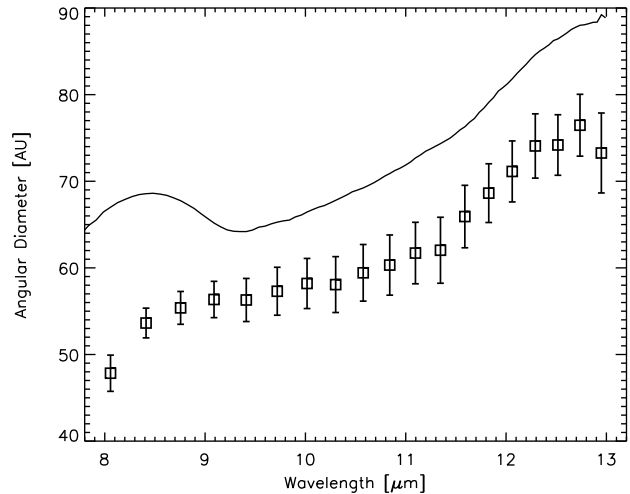}
\caption{The angular diameter of IRAS\,18006-3213 (solid line) is
  compared to the angular diameter of the dust emission around
  HD\,52961 (squares). The latter object is a binary post-AGB object
  surrounded by a circumbinary disc with a binary period of 1300\,days
  \citep{Deroo_2006}. Clearly both curves provide a remarkable
  similarity both in size and in spectral
  behaviour.}\label{fig:comparison_with_hd52961_iras18006}
\end{figure}

\object{IRAS\,18006-3213} shows remarkable similarities to the binary
post-AGB objects which are known to be surrounded by a
\emph{circumbinary} disc, since the detected orbits are all within the
dust sublimation radius \citep[for a recent review we refer
to][]{Vanwinckel_2003}. These objects have similar high luminosities
and it is shown that high processing degrees are observed for a very
high fraction of objects \citep[over 60\,\% of the N-band spectra show
a crystallinity fraction larger than
40\,\%,][]{DeRuyter_2006phd}. Also the visibility measurements of
\object{IRAS\,18006-3213} provide a very similar size and spectral
behaviour as observed for the only binary post-AGB object resolved with
MIDI \citep[i.e. \object{HD\,52961},][]{Deroo_2006}. This is shown in
Fig. \ref{fig:comparison_with_hd52961_iras18006}, where the absolute
angular diameter is given for both objects. The similarity between
both curves indicates that similar physical processes are active in
both objects, providing again evidence for the circumbinary nature of
the disc around \object{IRAS\,18006-3213}.  We do not expect to
see a similar phase jump for \object{HD\,52961} as for
\object{IRAS\,18006-3213} because the post-AGB photosphere of
\object{HD\,52961} is much hotter (and O-rich), making the
photospheric spectrum devoid of strong molecular lines.

\section{Conclusions}\label{sect:conclusions_iras18006}
The spectroscopic and interferometric data presented on
\object{IRAS\,18006-3213} provide clear and direct evidence that the
oxygen rich reservoir is located in a compact region close to the
star: 
\begin{itemize}
\item The stability of the N-band spectrum and the observation that
  the circumstellar environment consists of highly processed silicate
  grains both indicate the existence of a long-lived reservoir in the
  system where the processing occurred.
\item The interferometric observations show the very compact nature of
  the N-band emission, again providing strong evidence that this
  emission originates from a long-lived Keplerian disc.
\item The interferometric observations show an asymmetric N-band
  emission. This provides direct evidence of the non-spherical nature of
  the emission.
\end{itemize}

Using the spectroscopic and interferometric data, we show that the
most favourable geometry supported by the data is a long-lived
disc-like reservoir around the system which is seen under an
inclination. The formation of this disc is likely linked to the
strongly suspected binary nature of this object. We argued that
the bimodal appearance of the silicate N-band profile in silicate
J-type stars (processed versus non-processed grains), could be the
result of a different location of the dust in the systems which is
either circumcompanion (e.g. in \object{V778\,Cyg}) or circumbinary
(e.g. in \object{IRAS\.18006-3213}). The testing of this will require
direct detection of the orbits, which will need extensive radial
velocity monitoring campaigns and/or interferometric measurements
providing (closure) phase information.

%
%
%
%
\begin{acknowledgements}
The authors would like to thank the staff of the Paranal observatory
for observing the MIDI data.  P.~D., H.~V.~W., T.~V. and
M.~R. acknowledge financial support from the Fund for Scientific
Research of Flanders (FWO).
\end{acknowledgements}

\bibliographystyle{aa}

\end{document}